%Paper: hep-ph/9206213
%From: GOLDBERG@neuhep.cas.northeastern.edu
%Date: Mon, 8 Jun 92 17:05 EST

\input phyzzx
\def\ra{\rightarrow}
\def\prd#1{{\sl Phys.~Rev.}~{\bf D#1}\ }
\def\prb#1{{\sl Phys.~Rev.}~{\bf B#1}\ }

\def\plett#1{{\sl Phys.~Lett.}~{\bf #1B}\ }

\Pubnum={$\caps NUB - \the\pubnum $}
\def\abstract{\vskip\frontpageskip\centerline{\twelverm ABSTRACT}
              \vskip\headskip }
\def\author#1{\vskip 0.3cm
\titlestyle{\twelvecp#1}\nobreak}

\def\NEU{\address{Department of Physics\break
      Northeastern University, Boston, Massachusetts 02115}}
\def\us#1{\undertext{#1}}
\def\fract#1#2{\ifmmode{\textstyle{#1\over #2}}\else
    ${\textstyle {#1\over #2}}$\fi}
\def\quarter{\ifmmode{\textstyle{1\over 4}}\else ${\textstyle {1\over 4}}$\fi}
\def\half{\ifmmode{\textstyle{1\over 2}}\else ${\textstyle {1\over 2}}$\fi}
\def\third{\ifmmode{\textstyle{1\over 3}}\else ${\textstyle {1\over 3}}$\fi}

\def\ni#1{\noindent (#1)$\quad$}
\REF\cornwall{J.~M.~Cornwall, \plett{243},  271 (1990).}
\REF\hgs{H.~Goldberg, \plett{246}, 445 (1990).}
\REF\hgv{H.~Goldberg, \prd{45}, 2945 (1992).}
\REF\zakh{V.~I.~Zakharov, Max Planck preprint MPI-PAE-PTH-11-91, March
1991.}
\REF\lip{L.~N.~Lipatov, Sov.~Phys.~JETP {\bf 45}~(2), 216 (1977).}
\REF\knowb{It must be noted that we generally do not know the value
of $b.$ This is the case in the spontaneously broken electroweak
theory.}
\REF\hgmtv{H.~Goldberg and M.~T.~Vaughn, Northeastern preprint
NUB-3043/92-TH.}
\REF\endpt{The end-point problem with $8\ra 7$ is obvious:
many pairwise
additions in the maximal state $\ket{1,2,3,4,5,6,7,8}$ will give a momentum
already present in the spectator particles.}
\REF\voloshin{M.~B.~Voloshin, \prd {43} (1991) 1726. See also a recent
discussion by J.~M.~Cornwall and G.~Tiktopoulos, UCLA preprint
UCLA/91/TEP/55.}
\REF\bachas{C.~Bachas, Ecole Polytechnique preprint A089.1191,
December 1991.}
\REF\pdot{ I find for all cases  of interest,
$$\left({d\over dt}|A_N(t)|^2 \right)_{\rm max}
\sim |A_N(t)|^2_{\rm max}\ \ .$$ Thus, I will phrase the discussion
in terms of $\a1n .$}
\REF\parisi{G.~Parisi, \plett{68},117 (1977).}
\REF\pph{A.~Houghton, J.~S.Reeve, and D.~J.~Wallace, \prb{17}, 2956
(1978).}

\def\one8{1\Rightarrow 8}
\def\onen{1\rightarrow N}
\def\a1n{A_{\onen}}

\def\calp{{\cal P}}
\def\ncrit{N_{\rm crit}}
\def\kcrit{K_{\rm crit}}
\def\impi{{\rm Im}\ \Pi}
\def\g2{g^2}
\def\ebg{e^{-b/\g2}}
\def\anmax{|A_N|_{\rm max}}
\def\aemax{|A_8|_{\rm max}}
\def\anmaxb{|A_N^{\rm Born}|_{\rm max}}
\def\aemaxb{|A_8^{\rm Born}|_{\rm max}}

%%%%%%%%%%%%%%%%%%%%%%%% Title page %%%%%%%%%%%%%%%%%%%%%%%%%%%%%%%%%%%%
\Pubnum{\sl NUB-3044/92-Th}
\date{\sl April~1992}
\pubtype{}
\titlepage
\title{\bf Exact Nonperturbative
Unitary Amplitudes for $\onen$ Transitions }
\vskip .8cm
\author{Haim Goldberg}
\vskip .5cm
\NEU
\vskip .4cm

\abstract

I present an extension to arbitrary $N$ of a previously proposed
field theoretic model, in which unitary amplitudes for $1\ra 8$ processes were
obtained. The Born amplitude in this extension has the behavior
$\a1n^{tree}\ =\ g^{N-1}\ N!$ expected in a bosonic field theory.
Unitarity is violated when $|\a1n|>1,$ or when $N>\ncrit\simeq e/g.$
Numerical solutions of the coupled Schr\"odinger equations shows that
for weak coupling and a large range of
$N>\ncrit,$   the exact unitary amplitude is reasonably fit by a factorized
expression
$$|\a1n|_{\rm max} \
\simeq \ 0.73\ \cdot\ {1\over N}\ \cdot \exp{(-0.025/\g2)}\ \ .$$
The very small size of the coefficient $1/\g2$ , indicative of
a very weak exponential suppression, is not in accord with standard
discussions based on saddle point analysis, which give a coefficient of
$\sim 1.\ $ The weak dependence on $N$
could have experimental
implications in  theories where the exponential suppression is weak
(as in this model). Non-perturbative contributions to
few-point correlation functions in this theory would arise at order
$K\ \simeq\ \left((0.05/\g2)+ 2\ ln{N}\right)/ \ ln{(1/\g2)}$
in an expansion in powers of
$\g2.$

\vskip 1.5cm
\endpage
%%%%%%%%%%%%%%%%%%%%%%%%%%%%%%%%%%%%%%%%%%%%%%%%%%%%%%%%%%%%%%%%%%%%%%%%%

\frontpagefalse
\pagenumber=1

Tree-level calculations of amplitudes for scattering processes in which many
bosons are produced fail to obey unitarity when the number produced is too
large. Specifically, in either massive scalar [\cornwall,\hgs] or vector
[\hgv] field
theories, the $1\ra N$ tree amplitude for fixed angle scattering has
the unacceptable behavior
$$\a1n \sim g^{N-1} N!$$
for large $N,$ where $\g2$ is the quartic coupling. Cross sections then violate
unitarity when $N\gsim 1/\g2$ for $E,N$ large, $E/N$ finite. As shown
explicitly in refs.[2,3], the problem originates in the coherence of the
approximately $N!$ graphs which contribute. The cure surely lies in the
summation of all loops, since their  contribution is equal to that of
the tree graphs
precisely when the latter become large. The question remains: does the unitary
damping suppress the cross section at an exponentially small value? or do
coherence effects allow an experimentally interesting cross section for
producing many massive bosons in a high energy collision? Independent of the
phenomenological application lies the question as to how to calculate high
energy, multiparticle processes in field theory.

Zakharov [\zakh] has given a thoughtful argument for exponential suppression,
which I will  simply paraphrase. Consider the dispersion relation for
the fourier transform of
some two point function in field theory or quantum mechanics,with
enough subtractions to make it convergent. At zero energy, we have

$$\Pi(0)={1\over \pi}\int_m^{\infty} {dE\over E}\ \impi(E)\
\ , \eqn\disp$$
where
$$\impi(E)\sim \sum_N\ |A_{1\ra N}|^2\; \theta(E-mN)\ \
.\eqn\spect $$
Suppose we now {\it assume} that $\Pi(0)$ has an asymptotic expansion
in coupling
$$\Pi(0)=\sum_k\  a_k\ g^{2k}\ \ ,\eqn\asymp$$
where for large $K$
$$\left|\Pi(0)-\sum_1^K a_k\ g^{2k}\right|< (\g2/b)^K\ K!\ \ .$$
Then, by the usual arguments, the error in truncating the series is
minimized at $K=\kcrit = b/\g2,$ and the error in omitting
the remainder is bounded by
$$ \Delta\ \Pi(0)\  \lsim \ \ebg\  \  .\eqn\bound$$
At this point the discussion may take the following form:
\item{}Suppose we either {\it know } or {\it think} we know the value
of $b.$ (For example, in the scalar $\phi^4$ field theory, it is often
assumed that $b$ is given correctly as $\ 16\pi^2\ $ by the
saddle point analysis of Lipatov [\lip].) Then the high order behavior
(for $k\ge b/\g2)$
of $\impi$ is bounded by $\exp{(-b/\g2)}.$ Assuming that the  high
{\it multiplicity} $n\ge b/\g2$ piece of $\impi$ is an important
component
of the high {\it order} piece, one concludes that high multiplicity
contributions to $\impi$ (or the cross section, in field theories) are
bounded by [\knowb]
$$\sum_{N\ \gsim \ b/\g2}\ |A_{1\ra N}|^2\; \le\  \ebg\ \ .\eqn\csbd$$

One may
also turn the argument around: suppose that we have some evidence that
multiparticle contributions to $\impi$ are as large as
$\exp{(-c/\g2)}.$ Then, from the dispersion relation, we may conclude
that the perturbation series for $\Pi(0)$ becomes
untrustworthy for $k\gsim c/\g2.\ $ If $c\ll 1,$
the situation becomes interesting, on
two counts : (1) although {\it formally} exponentially
suppressed, there may be {\it experimentally}
interesting multiparticle cross sections  if $\g2$ is not too small
\ (2) as a consequence of the dispersion relation, the two point function
may have interesting non-perturbative contributions at relatively low
order $k\sim c/\g2.$

In this paper, I will propose an extension to arbitrary $N$ of a field
theoretic model [\hgmtv]
previously applied to a calculation of a unitary $1\ra 8$
transition  amplitude. In this extension, the (time-dependent)
amplitudes $A_{1\ra N}$
may be directly (numerically) calculated . The result of the present
calculation is that in the weak
coupling limit, the amplitude $\a1n$ is very weakly exponentially
suppressed:
$$\a1n\ \sim \ e^{-0.025/\g2}\ \ ,\eqn\weakexp$$
with the implications mentioned in the preceding paragraph and in the
abstract.

A field theoretic model, loosely based on a $\phi^3$ field theory, has been
presented [\hgmtv] as a
laboratory for the study of multiparticle amplitudes. The model
is defined through its Hamiltonian
$$H= \sum_{k=1}^\infty a^{\dagger}_k a_k + \half g\ \calp\sum_{j,k=1}^\infty
a^{\dagger}_{j+k}\ a_j\ a_k\ \calp + \ \ h.c.\ \ \ \  .\eqn\ham$$
The modes labeled by $i,j,k$ will be called ``momenta'', and the action of the
hermitean projection operator $\calp$ on a state vector $\ket{\psi}$ is as
follows:
$$\calp\ket{\psi}=0 $$
if there is more than one quantum in the state with any given momentum
and
$$\calp\ket{\psi}=\ket{\psi}$$
otherwise.
$\calp$ has been introduced in order to approximate
the infinite volume effect of
field theory (in box normalization): namely, one generally omits
consideration of
amplitudes for transitions to states with more than one particle
in a given (discrete) momentum state. Thus, we exclude ``laser'' effects. In
this sense, there is an exclusion principle without imposing
anticommutation relations and antisymmetrization.
Other than that, $H$ resembles a $\phi^3$ field theory in a cavity, with no
temporal or spatial derivatives in the lagrangian. It is also a kind of matrix
model.

It is an important consequence of \ham\ that the momentum operator
$$P=\sum_{k=1}^{\infty}\ k\ a^{\dagger}_ka_k\ \ \ \eqn\mom $$
is a constant of the motion. Thus, the Hilbert space factorizes into subspaces
with definite $P.$ Because of the positivity of all of the momenta, {\it these
will be finite dimensional subspaces.} Within a subspace of given $P,$
there will be sectors characterized by different numbers of particles
$n.$ A subspace with definite $P$ may contain a {\us{maximal}} state,
namely an $N$-particle state with momenta $k=1,2,\ldots N,\ $such that
$P=N(N+1)/2.$ I will restrict
my study to these subspaces, which can be labeled by $N.$
For $N=8,$ the number of states with $n=1,2,\ldots 8$ particles is
(1,\ 17,\ 91,\ 206,\ 221,\ 110,\ 21,\ 1) respectively.
A numerical analysis of this
case [\hgmtv] proceeded as follows: with the system initially in the
one-particle state, the 668 coupled (complex) time-dependent
Schr\"odinger equations were numerically integrated in order to obtain
the maximum 8-particle amplitude $|A_8(t)|_{\rm max}.$ The result
indicated an approximate behavior for the $1\ra 8$ transition
$$\aemax \simeq e^{-0.20/\g2}$$
for $\g2>0.07.$ This is the value of $\g2$ for which the Born amplitude
violates unitarity: $|A_8^{\rm Born}(t)|_{\rm max}\ge 1.$

In general, the total number of states for a given $P$ can be obtained as the
exponent of $x^P$ in the expansion of the generating function
$$\prod_{j=1}^\infty (1+x^j)= \sum_{P=0}^\infty {\cal N}_P\ x^P\ \ \
.\eqn\gen$$
This reveals that even an extension to $N=9$ means quadrupling the
subspace from 668 to 2048 states. Thus, exact numerical analysis for
larger $N$ rapidly becomes impractical.

How then to extend the results beyond $N=8?$ Clearly,
an approximation to the dynamics of the model at large $N$
is required; however, it is imperative that the extension to arbitrary $N$
$(a)$ retain the unitarity property and $(b)$ display the
$g^{N-1}\ N!$ behavior in
the tree, or Born approximation.

A detailed study of the coupling structure of the Hamiltonian (7) for $N=8$
yields a strong clue as to how this extension may be accomplished. If one asks,
how many states (on the average) with $n-1$ particles does a particular state
with $n$ particles couple to, then one arrives at the following list:
\settabs 3\columns
\+&$8\ra 7:\ $ 16 states\cr
\+&$7\ra 6:\ $ 15$\pm$ 1 states\cr
\+&$6\ra 5:\ $ 12$\pm$ 1 states\cr
\+&$5\ra 4:\ $  9$\pm$ 1 states\cr
\+&$4\ra 3:\ $  6        states\cr
\+&$3\ra 2:\ $  3 states\cr
\+&$2\ra 1:\ $  1 state\cr

If we ignore the $8\ra 7$ result, and are liberal with the $7\ra 6,$ then it is
not too far amiss to conclude that, approximately, each state with $n$
particles
couples to an average of $n(n-1)/2$ states with $n-1$ particles.\
[\endpt]
This appeals
to intuition: adding each (unequal) pair of `momenta' in the given state with
$n$ particles will, in large probability, give one of the possible states with
$n-1$ particles.
Thus, I propose as an extension to \ham\
a quantum mechanical system in which there
are simply $N$ states labelled by $n=1,\ldots N,$ differing in energy
by a constant amount (taken to be 1 in our units), and in which each
state $\ket{n}$ is coupled with strength $g\ n(n-1)/2$ to the state
$\ket{n-1}.$ In the interaction representation the dynamics of this
model is embodied in the system of time dependent Schr\"odinger equations
$$ \eqalign{ i \dot A_n(t)\ =\ & g\
    \left({n(n-1)\over 2}\ A_{n-1}(t)\ e^{it} \right.\ +  \left.
{n(n+1)\over 2}\ A_{n+1}(t)\ e^{-it}\right)\crr
& n=1\ldots N\crr}\eqn\schr$$
where $A_0(t)=A_{N+1}(t)\equiv 0.$ For a given $N,$ these equations
may be solved by first diagonalizing the $N\times N$ Hamiltonian, or
by direct numerical integration. But first, I state
without details of proof a result
which establishes this system as a viable laboratory in which to study
the large-$N$ problem:

\smallskip
If $A_n(0)=\delta_{n1},$ then in Born approximation
$$A_n^{\rm Born}(t) = g^{n-1}\ \left({1-e^{it}\over 2}\right)^{n-1}\ n!\ \
.\eqn\bornt$$
This can be proven by direct subsitution into Eq.\schr\ with no rescattering
(\ie,\ dropping the second term on the right). It was actually arrived
at constructively by using Laplace transforms.
\smallskip

As a result of \bornt,\
$$\anmaxb \ = \ g^{N-1}\ N!\ \ .\eqn\anbmax$$
I note that in the exact version of the model, with $N=8,$ [\hgmtv] it was
found that
$$\aemaxb \ \simeq 0.27 \ g^7\ 8!\ \ .\eqn\aebmax$$
Note also that in  Born approximation  the anharmonic oscillator
with coupling $\quarter \g2 x^4$ gives [\voloshin,\bachas]
$$\anmaxb\ \sim \  g^{N-1}\sqrt{ N!}\ \ .\eqn\ananh$$
In both the present case and in the case of the anharmonic oscillator
the state $\ket{N}$ with unperturbed energy $N$ is normalized to
unity.

Since there is a normalization condition $\sum_n|A_n(t)|^2=1,$ it is
seen from Eq.\anbmax\  that unitarity will certainly be violated in Born
approximation when
$$N\gsim \ncrit = {e\over g}\ \  .\eqn\encrit$$
For a given $\g2,$ this defines the value of $N$ for which
rescattering terms are essential.

It is then a simple matter to go back to Eq. \schr,\ and find
(numerically) the maximum
value $\anmax$ attained by $A_N(t)$ as a function of $N,\ \g2.$
[\pdot]
\bigskip
\noindent{\us{Results}:} The results of this work are encompassed in
Figs. (1) and (2). In Fig. (1),  I display on a $log-log$ plot the $N$
behavior of $\anmax,$  for a large range of values of $\g2.$ The range
of $N$ for each $\g2$ is chosen to lie above $\ncrit.$ Two striking
observations may be made from the graph: (1) The $N$-behavior is
universal over the whole array of $\g2\  $\dash\  the amplitude {\it
factorizes};
(2) The curves are excellently fit with a simple inverse
proportionality $\anmax\ \sim \ 1/N.\ $ The wavy oscillations in the
curves for the larger values of $\g2$ are real: they can be seen in
detail in Fig. (3) as an odd-even effect.

In Fig. (2), I plot $\anmax$ {\it vs.} $1/\g2$ on a $log$ plot
for a range of values of $N.$ Again, one notes the factorizability.
For the smallest
values of $\g2,\ $the curves are well fit by the exponential form
$\anmax\ \sim \exp{(-0.025/\g2)}.$ In sum, therefore, the
results of the numerical study of the equations \schr\ is that
$$|\a1n|_{\rm max}\ \simeq \
0.73\ \cdot\ {1\over N}\ \cdot \exp{(-0.025/\g2)}\ \
.\eqn\param$$
I close with some discussion of these results.
\medskip
\noindent{\us{Remarks and Conclusions:}}
\ni{1} First, the factorizability. The results of Lipatov [\lip] and
the graphical analysis of Parisi [\parisi] support the notion that the
contribution of a high order $K$ of perturbation theory to an
$N$-legged Green's function should be (roughly) independent of $N,$
for $N$ not too large.
What I find here is that the functional dependence on coupling
constant of the
exact non-perturbative $N$-point function is  independent of $N,$ in
the region where $N$ {\it is} large $(N>e/g.)$

\ni{2} The behavior of $\a1n$ in the model being examined is totally
different from its behavior in the case of the anharmonic oscillator.
In that case, for $Ng^2$ not too large, $\anmax\ \sim \
\exp{(-N)}.\ $ [\voloshin,\bachas]

\ni{3}Another surprising result is the generation of a small
dimensionless number, namely 0.025, the coefficient of $1/\g2$ in the
exponential suppression. Typical saddle point analyses (even in the
$\phi^3$ theory [\pph]) give coefficients of $O(1)$ ({\it modulo}
factors of $(4\pi)^{d-2}$ for $d$ dimensions). The implication is that
for $N >\ncrit,$ the theory becomes strongly coupled for very small
values of $g^2.$ The next point is
related to this one.

\ni{4} The contribution of the $1\ra N$ excitation
to a few-point function (such as the $1\ra 1$ transition amplitude)
can be estimated as
$$\VEV{1,\ t=T|1,\ t=0}_N\ \sim \ \anmax^2\ \ \ .\eqn\oneone$$
If we wish to see at what order $K$ the
non-perturbative contribution of the right-handed side competes with a
perturbative development of the left-hand side, we can use the
parameterization \param\  and set
$$g^{2\kcrit}\ \simeq \ \exp{(-0.05/\g2)}/ N^2\ \ \ ,\eqn\kcr$$
with $N\ge e/g.$ This gives the formula in the abstract. Even for
$g^2=0.005$ and $N=100,$  this gives $\kcrit=3.6,$ a very low value.

\ni{3} Finally, a comment on the limitations of the numerical study. As
$N$ becomes large and/or $\g2$ becomes small, the roundoff errors become
more important. At some point, $\anmax^2$ is smaller than the
deviation from unitarity  $\delta=\left|\sum_n|A_n(t)|^2\ - 1\right|$
caused by finite numerical accuracy. This occurred for
$\g2< 0.005$ (and $N>\ncrit).$
The results for such small values of coupling may
therefore be untrustworthy, and at present I am examining alternate
methods of exploring the full range of $\g2$ in the non-perturbative region.

To conclude, I have found in a unitary model a large $\onen$ amplitude
which is only very weakly exponentially suppressed. The dependence on
$N$ is also very weak. This may hold out some possibility for
observing large multiplicity central  events in high energy
collisions.
\vskip  2cm

\centerline{\bf Acknowledgements}
\vskip 1cm
  The author would
like to express appreciation to the participants of the Yale-Texas
Workshop on Electroweak Baryon Number Violation for stimulating
discussion following a talk based on this work. This research  was
supported in part by the National Science Foundation under
Grant No. PHY-9001439, and by the Texas National Laboratory Research
Commission under Grant No. RGFY9114.

\vskip 2cm
\centerline {\bf Figure Captions}
\vskip 1cm
\item{Fig.\ 1}$\quad$ The maximum value attained by $|A_N(t)|$ as a
function of $N,$ for various values of $\g2.$ The Born approximation
fails for $N\ge\ncrit\ = e/g.$ For small
$\g2, $ large $N,$\ $\anmax\ \sim \ 1/N.$

\item{Fig.\ 2}$\quad$ $|\a1n|_{\rm max}$ {\it vs.} $1/\g2$ for three values
of $N.$ The behavior in $1/\g2$ is independent of $N:$ $\a1n\sim
\exp{(-0.025/\g2)}.$

\item{Fig.\ 3}$\quad$ Detail of Fig. (1) for $g^2\ =\ 0.10,$ showing
origin of waviness as an odd-even effect.
\refout
\bye